\pdfoutput=1
\documentclass[11pt]{article}

\usepackage[final]{acl}

\usepackage{times}
\usepackage{latexsym}

\usepackage{amsmath}
\usepackage{amssymb}
\usepackage{mathtools}
\usepackage{amsthm}
\usepackage{bm}
\usepackage{url}

\usepackage[T1]{fontenc}
\usepackage{color}
\usepackage{colortbl}
\usepackage{xcolor}

\usepackage[utf8]{inputenc}
\usepackage{arydshln}
\usepackage{booktabs}

\usepackage{multirow}
\usepackage{microtype}

\usepackage{inconsolata}
\usepackage{paralist}

\usepackage{graphicx}
\usepackage{tcolorbox}
\usepackage{adjustbox}
\usepackage{xspace}

\definecolor{lightyellow}{RGB}{247,242,196}

\definecolor{deepgreen}{RGB}{28,198,139}
\definecolor{deepblue}{RGB}{18,113,203}
\definecolor{deepred}{RGB}{252,75,77}

\definecolor{lightgray}{RGB}{219,219,219}
\definecolor{deepgray}{RGB}{193,193,193}
\definecolor{nmgray}{RGB}{229,229,229}
\definecolor{nmblue}{RGB}{216,234,247}

\newcommand{\textlogo}{\raisebox{-4pt}{\includegraphics[width=1.8em]{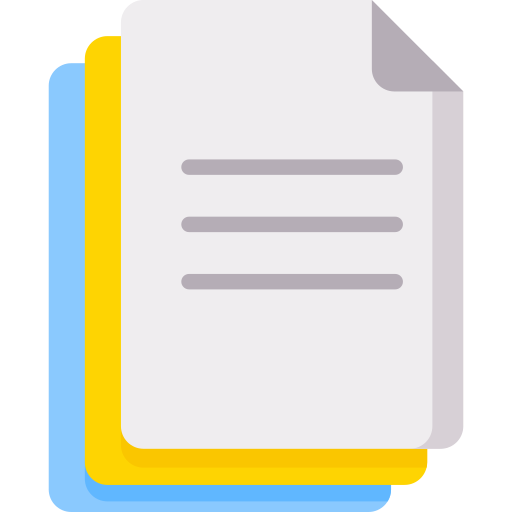}}\xspace\xspace}

\newcommand{\videologo}{\raisebox{-4pt}{\includegraphics[width=1.8em]{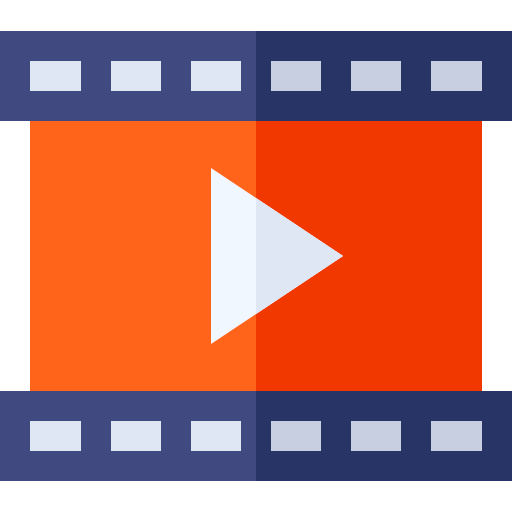}}\xspace\xspace}

\newcommand{\imagelogo}{\raisebox{-4pt}{\includegraphics[width=1.8em]{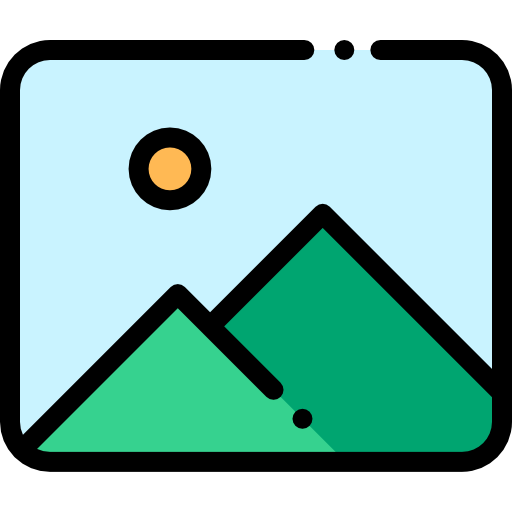}}\xspace\xspace}

\newcommand{\audiologo}{\raisebox{-4pt}{\includegraphics[width=1.8em]{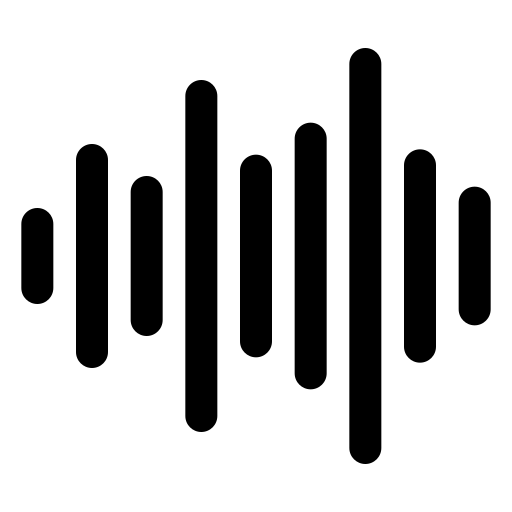}}\xspace\xspace}

\title{\emph{Recognizing Everything from All Modalities at Once}:\\ Grounded Multimodal Universal Information Extraction}

\author{
Meishan Zhang\textsuperscript{\rm 1}, Hao Fei\textsuperscript{\rm 2}\Thanks{ Corresponding author}, Bin Wang\textsuperscript{\rm 1}, Shengqiong Wu\textsuperscript{\rm 2}, Yixin Cao\textsuperscript{\rm 3}, Fei Li\textsuperscript{\rm 4}, Min Zhang\textsuperscript{\rm 1}  \\
\textsuperscript{\rm 1} Harbin Institute of Technology (Shenzhen),  \,  \textsuperscript{\rm 2} National University of Singapore,\\
\textsuperscript{\rm 3} School of Computer Science, Fudan University,  \,  \textsuperscript{\rm 4} Wuhan University \\
\texttt{zhangmeishan@hit.edu.cn,\, haofei37@nus.edu.sg,\, 23s051047@stu.hit.edu.cn,\, swu@u.nus.edu}\\
\texttt{caoyixin2011@gmail.com,\, lifei\_csnlp@whu.edu.cn,\, zhangmin2021@hit.edu.cn}
}

\begin{document}
\maketitle

\begin{abstract}
In the field of information extraction (IE), tasks across a wide range of modalities and their combinations have been traditionally studied in isolation, leaving a gap in deeply recognizing and analyzing cross-modal information. 
To address this, this work for the first time introduces the concept of \emph{grounded Multimodal Universal Information Extraction} (\textbf{MUIE}), providing a unified task framework to analyze any IE tasks over various modalities, along with their fine-grained groundings. 
To tackle MUIE, we tailor a multimodal large language model (MLLM), \textsc{Reamo}, capable of extracting and grounding information from all modalities, i.e., `\emph{recognizing everything from all modalities at once}'.
\textsc{Reamo} is updated via varied tuning strategies, equipping it with powerful capabilities for information recognition and fine-grained multimodal grounding.
To address the absence of a suitable benchmark for grounded MUIE, we curate a high-quality, diverse, and challenging test set, which encompasses IE tasks across 9 common modality combinations with the corresponding multimodal groundings.
The extensive comparison of \textsc{Reamo} with existing MLLMs integrated into pipeline approaches demonstrates its advantages across all evaluation dimensions, establishing a strong benchmark for the follow-up research.
Our resources are publicly released at \url{https://haofei.vip/MUIE}.
\end{abstract}

\section{Introduction}

IE is a pivotal topic \cite{Li00WZTJL22}, encompassing subtasks such as Named Entity Recognition \citep[NER; ][]{nadeau2007survey}, Relation Extraction \citep[RE; ][]{miwa2016end}, and Event Extraction \citep[EE; ][]{ahn2006stages}, which plays a crucial role in constructing domain-specific knowledge bases \cite{bosselut2019comet} and in facilitating deep semantic understanding of data \cite{CASIE}. 
In reality, a vast amount of information is conveyed through modalities beyond text. 
Consequently, research in IE has evolved from focusing solely on textual data to embracing various other modalities, leading to the development of multimodal IE \citep[MIE; ][]{liu2019graph}, e.g., images, videos, and audio. 
Despite the growing research efforts dedicated to MIE, the exploration in this area remains insufficiently developed.
We argue that several critical aspects must be fully considered for future MIE research trends.

\begin{figure}[!t]
\centering
\includegraphics[width=0.98\columnwidth]{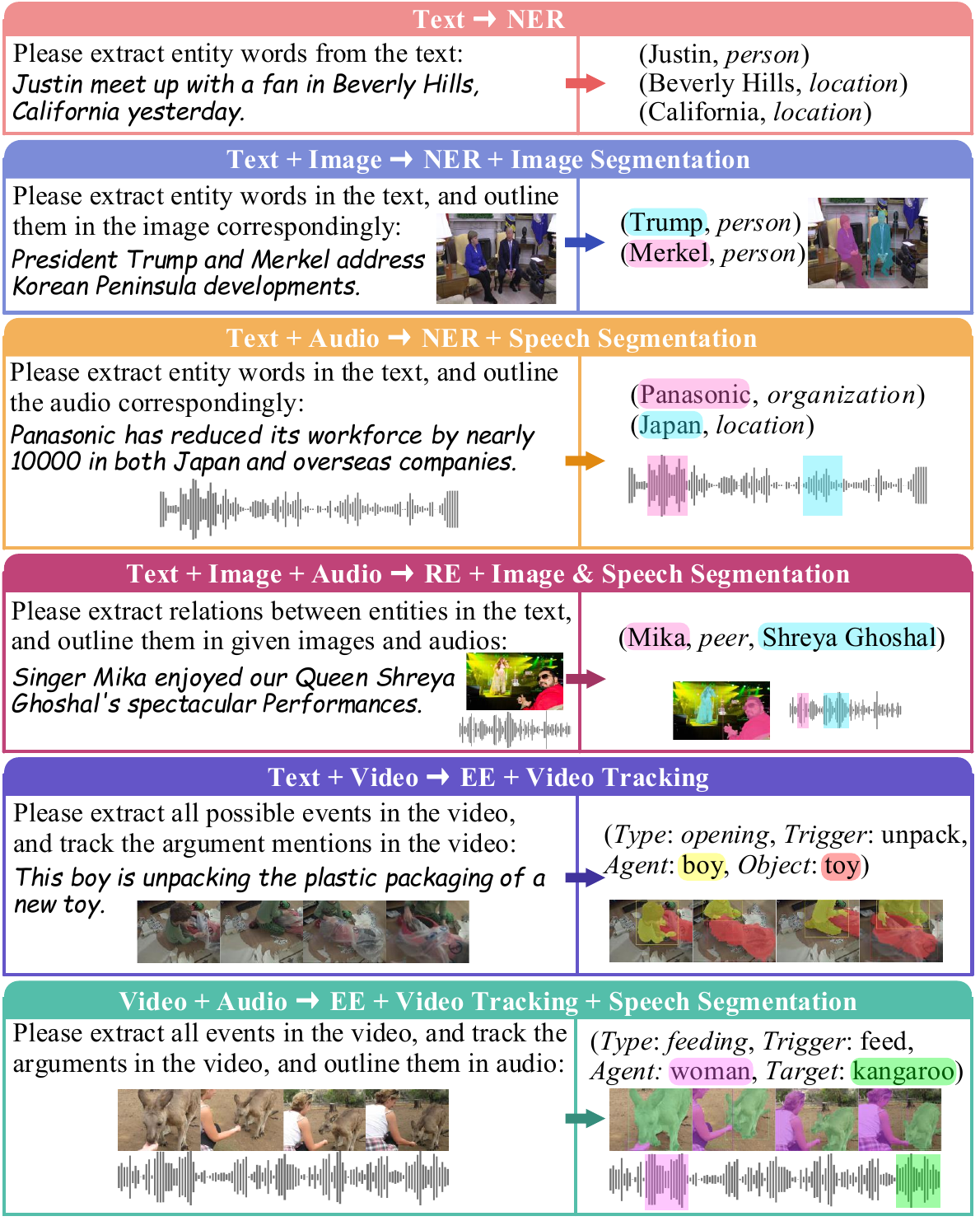}
\caption{
Examples of grounded multimodal universal information extraction (MUIE).
}
\label{fig:intro}
\vspace{-4mm}
\end{figure}

\textbf{Firstly}, current studies primarily investigate MIE tasks within individual modalities (or certain modality combinations) \cite{sun2021rpbert,chen2022good}. 
With the existence of several modality categories and diverse definitions for different IE tasks, studying each modality separately to construct specialized MIE models would inevitably lead to resource wastage and inefficiency. 
In real-world applications, there is a constant need for building unified systems with ``one-for-all'' robust generalizability for faster practical deployment.
In light of the recent success of textual universal IE \citep[UIE; ][]{lu2022unified,FeiLasuieNIPS22}, MIE unification should also be promising.
\textbf{Second}, the majority of existing studies \cite{zhang2017improving} exhibit a bias toward text-centric IE outputs, necessitating the decoding of detailed textual IE labels and inherently prioritizing text.
While they often treat other modalities as auxiliaries and do not produce outputs for them, this practice does not align with reality, because all modalities can equally carry important information.
For example, even infants who have not yet learned to speak can learn about entities from vision. 
Thus, each modality should be treated equally, and detect fine-grained information from all given modalities.
\textbf{Last}, most current MIE \cite{zheng2021multimodal} involving multiple modalities (e.g., Image\&Text) tends to extract the modality-aligned part of the information under the assumption that different modalities associate with each other. 
However, in practical scenarios, the information carried by different modalities can be either shared \cite{li2022clip}, or unrelated \cite{wu-etal-2023-information}. 
This suggests that information should be flexibly recognized from any modality sources.

In response to these challenges, this paper is dedicated to pioneering a novel task, grounded Multimodal Universal Information Extraction (\textbf{MUIE}).
As illustrated in Fig. \ref{fig:intro}, MUIE aims to unify the modeling of various IE tasks (e.g., NER, RE, EE) with any (or combination) inputs across the most common modalities (e.g., text, audio, image, and video), and produce fine-grained multimodally grounded IE results. 
To solve MUIE, we consider taking advantage of the existing generative LLMs \cite{chatgpt,abs-2210-11416,vicuna} with in-context instructions \cite{dong2022survey}. 
We develop a novel multimodal LLM, \textsc{Reamo}, achieving ``\emph{\textbf{R}ecognizing \textbf{E}verything from \textbf{A}ll \textbf{M}odalities at \textbf{O}nce}''.
\textsc{Reamo} not only outputs all possible textual IE labels but also identifies corresponding groundings across other modalities: 
1) statically, by segmenting visual objects and audio speeches, and 2) dynamically, by tracking textual or vocal events in videos. 
Technically, \textsc{Reamo} employs a Vicuna \cite{vicuna} LLM as its core semantic reasoner, utilizing ImageBind \cite{abs-2305-05665} as a multimodal encoder to project image, video and audio inputs into LLM-understandable signals. 
At the decoding end, we integrate the SEEM \cite{zou2023segment} for visual grounding\&tracking, and the SHAS \cite{TsiamasGFC22} for audio segmentation, where the messages are passing from LLM to decoders through structured meta-response effectively.
Given input multimodal information, \textsc{Reamo} is able to output UIE label tokens as well as fine-grained groundings recurrently.

We then design learning objectives to tune \textsc{Reamo} to endow it with robust MUIE and cross-modal grounding capabilities. 
First, we repurpose existing textual UIE annotation into instruction format, and use it to tune the backbone LLM for activating the UIE ability. 
Then, we perform both coarse-grained instance-level and fine-grained grounding-aware cross-modal alignment learning, enhancing the \textsc{Reamo}'s capability in fine-grained multimodal semantic understanding.
Furthermore, we instruction-tune \textsc{Reamo} on specific corpus, to build its working behavior of generating structured meta-response texts.

In response to the absence of standard evaluation data for grounded MUIE, we further introduce an evaluation benchmark, where we annotate a high-quality testing set of 3,000 instances covering NER, RE, and EE tasks under 9 common modality combinations. 
The data further advances by annotating both modality-shared/-specific content to simulate aligned and misaligned modality scenarios.
Extensive zero-shot experiments on these benchmarks demonstrate that \textsc{Reamo} shows strong performance over existing MLLMs with respect to IE tasks and multimodal grounding.

Overall, we make three key contributions:
\begin{compactitem}
    \item To our knowledge, this is the first to propose a grounded MUIE setting, unifying all IE tasks across modalities, further with fine-grained multimodally grounded targets.

    \item We introduce an MLLM for the task, \textsc{Reamo}, excelling in MUIE prediction and achieving cross-modal grounding of static objects and dynamic events.

    \item We contribute a high-quality, diverse, and challenging dataset, setting an evaluation benchmark for follow-up grounded MUIE research.

\end{compactitem}

\begin{figure*}[!t]
\centering
\includegraphics[width=0.98\textwidth]{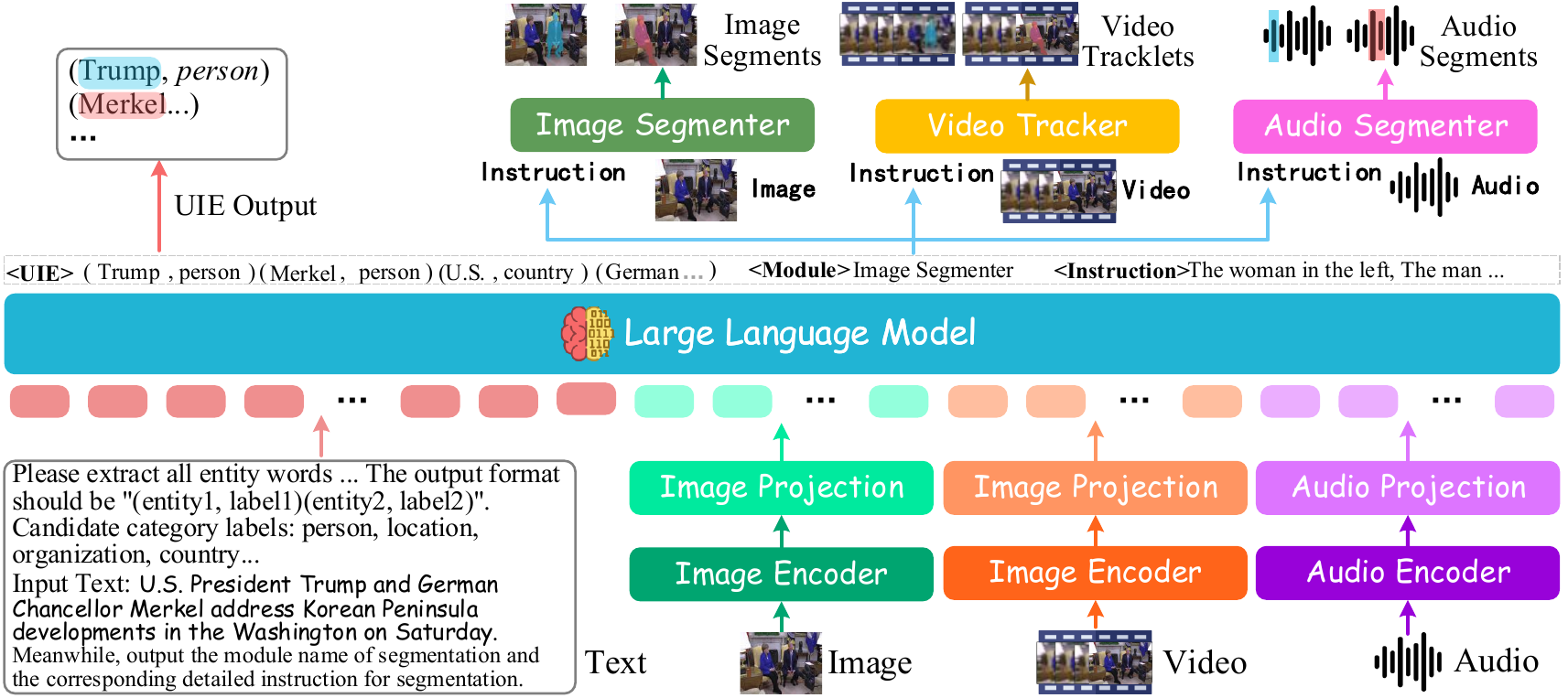}
\vspace{-1mm}
\caption{
An overview of the proposed \textsc{Reamo} MLLM architecture for grounded MUIE.
}
\label{fig:architecture}
\vspace{-4mm}
\end{figure*}

\section{Related Works}

\vspace{-1mm}

IE \cite{lample2016neural,wei2020novel,fei2021better,li-etal-2021-mrn,wang-etal-2022-entity,cao-etal-2022-oneee} has long been a significant research direction, consistently attracting substantial interest and focus for decades. 
As the world contains various modalities of information, MIE has been consequently introduced, e.g., multimodal NER \cite{sun2021rpbert,yu2023grounded,wang2023fine}, multimodal RE \cite{chen2022good}, and multimodal EE \cite{li2020cross}. 
However, the majority of existing well-established benchmarks for MIE still predominantly focus on texts, supplemented with images \cite{zhang2017improving,wu-etal-2023-information}.

Historically, IE research has treated different tasks as separate studies for a long \cite{sun2021rpbert,chen2022good,liu2024tkdp,zhang2024context}. 
Recently, \citet{lu2022unified} pioneer UIE, proposing to unify all IE tasks under a single generative model to produce all IE results, significantly reducing the maintenance cost for individual tasks.
With the latest rapid development of LLMs, the latest advancements have utilized LLMs with in-context prompting for UIE, achieving promising zero-shot performance. 
Similarly, the swift progress of MLLMs \cite{fei-etal-2023-scene} should also ignite hope for MUIE. 
Yet research in MUIE remains under-explored.

To our knowledge, the work most closely related to this paper is by \citet{sun2023multimodal}, who leverage existing MLLMs to unify various MIE tasks through a two-stage process of span extraction and classification in a multimodal QA format. 
Yet, we identify clear limitations in their approach that fall short of achieving comprehensive MIE unification.
Firstly, while their work only considers text and image modalities, lacking comprehensiveness, we broadly cover the four most common modalities.
Secondly, beyond addressing these limitations, we also introduce a novel MLLM tailored for grounded MUIE and contribute a new benchmark dataset for MUIE research. 
These efforts aim to pioneer the next stage of MUIE research.

\section{Task Definition: Grounded Multimodal Universal Information Extraction}

\vspace{-1mm}

We now give a formal definition of grounded MUIE.
Suppose the inputs are any of a text $T$, an image $I$, an audio $A$, a video $V$, or their combination.

\noindent\textbf{NER} task seeks to predict all possible textual labels of entities $\{E^{\text{ner}}\}$, with pre-defined entity types $C^{\text{ner}}\in \mathcal{C}^{\text{ner}}$ (e.g., person, location and organization), where each $E$ may correspond to a span within $T$, or visual region within $I$, or a speech segment within $A$.
We denote the visual grounding mask as $M_{img}$ and the speech segment as $M_{aud}$.

\noindent\textbf{RE} task aims to first identify all possible entities $\{E^{\text{re}}\}$ following the NER step, and then determine a pre-defined relation label $R^{\text{RE}}\in \mathcal{R}^{\text{RE}}$ for two entities $<E^{\text{RE}}_i, E^{\text{RE}}_j>$ that should be paired.
Also $E^{\text{RE}}$ should correspond to $T$, $I$, or $A$, as in NER.

\noindent\textbf{EE} task detects all possible structured event records that consist of event trigger $E^{\text{et}}$, event type $C^{\text{et}}\in \mathcal{C}^{\text{et}}$, event argument $E^{\text{ea}}$ and event argument role $C^{\text{er}} \in \mathcal{C}^{\text{er}}$.
Here $E^{\text{et}}$ and $E^{\text{ea}}$ correspond to a continuous span within $T$ or a speech segment within $A$.
Also $E^{\text{ea}}$ might refer to the visual region within $I$,
or the temporal dynamic tracklet in video $V$ (i.e., object tracking).
We denote the video tracking mask as $M_{vid}$.
$\mathcal{C}^{\text{et}}$ and $\mathcal{C}^{\text{er}}$ are pre-defined label sets.

Following \citet{wang2023instructuie}, we employ in-context learning \citep[ICL; ][]{dong2022survey} to prompt LLMs for MUIE, with the specific task executed depending on the user's intention.
In the bottom left of Fig. \ref{fig:architecture} we simply illustrate the ICL prompt.

\section{Our Proposed Model}

\subsection{MLLM Framework of \textsc{Reamo}}

Fig. \ref{fig:architecture} presents a schematic overview of \textsc{Reamo} MLLM, which consists of three main parts: multimodal encoder, LLM, and decoder for UIE prediction \& multimodal grounding.

\vspace{-1mm}
\paragraph{Multimodal Encoding.}
While \textsc{Reamo} takes four types of modality sources, except texts that are directly input to LLM, the image, audio, and video inputs should be encoded.
Following \citet{wu2023next,fei2024video,fei2024enhancing}, we leverage the high-performance ImageBind \cite{abs-2305-05665} as a unified multimodal encoder.
Then, via a projection layer, different input representations are aligned into language-like embeddings that are understandable to the LLM.

\vspace{-1mm}
\paragraph{LLM Reasoner.}
An LLM serves as the center unit of \textsc{Reamo} for content semantics understanding and reasoning.
Specifically, we follow the most common practice, using the Vicuna-v1.5 \cite{vicuna} as the backbone LLM.

\begin{tcolorbox}[fontupper=\linespread{0.9}\selectfont,]
{
\small
$\blacktriangleright$ \textbf{\normalsize\color{red}Input:}\\

\textbf{<Text to analyze>}\\
\textlogo\\
\textbf{<Non-text Modality>}\\
\imagelogo | \audiologo | \videologo\\
\textbf{<Prompt>}\\
\emph{Please extract all entity words $\cdots$}\\

$\blacktriangleright$ \textbf{\normalsize\color{green}meta-response:}\\

\textbf{<UIE>}\\
(Trump, person) \\
(Merkel, person) \\
$\cdots$ \\
\textbf{<Module>}\\
Image Segmenter \\
\textbf{<Instruction>}\\
Segmentation: `\emph{A person}' \\
\vspace{-3mm}
}
\end{tcolorbox}

\vspace{-1mm}
\paragraph{MUIE Decoding with Grounding.}

The central LLM takes on the crucial role of decision-making. 
Based on the input prompt, LLM will produce textual responses, containing the UIE task results as well as the meta-response that will be used to call the downstream modules to generate fine-grained multimodal groundings.
Here, consider integrating the existing high-performance SEEM model \cite{zou2023segment} for image segmentation and video tracking \cite{OMGSeg,wu2024tokenization,fei2024vitron}, and the SHAS model \cite{TsiamasGFC22} for audio segmentation.
Specifically, the meta-response includes three parts, as shown above.
The corresponding `<Module>' and `<Instruction>' information will be passed to the corresponding modality segmenter(s), to activate them to generate grounding(s).

\subsection{MUIE Fine-tuning for \textsc{Reamo}}

With \textsc{Reamo} at hand, we now consider fine-tuning it through multiple objectives to enable \textsc{Reamo} with strong MUIE capability.

\vspace{-2mm}
\paragraph{UIE Instruction Tuning.}
Our initial goal is to equip the system with the fundamental capability for UIE in the text modality. 
To achieve this, we consider tuning the backbone LLM specifically for UIE. 
Following the practices of \citet{wang2023instructuie}, we repurpose existing annotation data to form a set of instruction-tuning datasets for UIE. 
To avoid the huge cost of fully updating the LLM, we leverage the LoRA technique \cite{HuSWALWWC22}, achieving the goal by tuning only a small subset of parameters, without altering the overall LLM.

\vspace{-2mm}
\paragraph{Multimodal Alignment Learning.}

\textsc{Reamo} integrates the ImageBind encoder to enable the LLM to comprehend basic multimodal signals. 
Following this, we proceed with multimodal alignment learning. 
We consider using the language-centric LLM as the core, requiring only the alignment of other modalities to text.
We mainly utilize the vast array of available `X-caption' pair data (`X' stands for image, audio, or video). 
We adopt an `X-to-text' generation, where the input is `X', and the LLM generates the corresponding caption. 
During this process, we fix the ImageBind and LLM while only updating the projection layer.

\vspace{-2mm}
\paragraph{Fine-grained Cross-modal Grounding-aware Tuning.}

Above we merely enable \textsc{Reamo} with a coarse-grained multimodal understanding. 
Yet our goal is to attain a subtle modality comprehension. 
Thus, we further engage in fine-grained multimodal grounding. 
Our primary approach revolves around utilizing existing `X-to-text' phrase grounding data, e.g., MS-COCO \cite{lin2014microsoft}, where \textsc{Reamo}'s input consists of textual phrases and grounded regional modality features, and then prompt LLM to determine their match.

% deepblue
\begin{table*}
 \centering
\fontsize{9}{12}\selectfont
\setlength{\tabcolsep}{2.5mm}
\begin{tabular}{lccc}
\hline
\multicolumn{1}{c}{\multirow{2}{*}{\textbf{Modality}}}
 & \multicolumn{3}{c}{ \textbf{Tasks}} \\
\cmidrule(r){2-4}
 & \textbf{NER} & \textbf{RE} & \textbf{EE}\\
\hline
\bf I & PASCAL-C \cite{mottaghi2014role} &	VRD \cite{lu2016visual} &	imSitu \cite{yatskar2016situation} \\
\bf V & \cellcolor{lightgray}  & \cellcolor{lightgray} & VidSitu \cite{Sadhu_2021_CVPR}  \\
\bf A & \cellcolor{lightyellow} ACE05\textcolor{deepred}{-Aud} \cite{walker2011ace} & ReTACRED \cite{wu2022towards} & \cellcolor{lightgray} \\
\hdashline
\bf T+I & Twt17 \cite{lu2018visual} & MNRE \cite{zheng2021multimodal} & M$^2$E$^2$ \cite{li2020cross}  \\
\bf T+V& \cellcolor{lightgray}  & \cellcolor{lightgray} &\cellcolor{lightyellow} VidSitu\textcolor{deepgreen}{-Txt}  \cite{Sadhu_2021_CVPR} \\
\bf T+A	& \cellcolor{lightyellow} ACE05\textcolor{deepred}{-Aud}  \cite{walker2011ace} &ReTACRED \cite{wu2022towards} & \cellcolor{lightgray}  \\
\hdashline
\bf I+A	& \cellcolor{lightgray} &\cellcolor{lightyellow} MNRE\textcolor{deepred}{-Aud} \cite{zheng2021multimodal} &\cellcolor{lightgray}  \\
\bf T+I+A	& \cellcolor{lightyellow} Twt17\textcolor{deepred}{-Aud} \cite{lu2018visual} &\cellcolor{lightgray} &\cellcolor{lightgray}  \\
\bf V+A	& \cellcolor{lightgray} & \cellcolor{lightgray} &\cellcolor{lightyellow} VidSitu\textcolor{deepred}{-Aud} \cite{Sadhu_2021_CVPR}  \\
\hline
\end{tabular}
% }
\vspace{-2mm}
\caption{
Summary of the grounded MUIE test data.
Items in the light yellow background mean they are the data after preprocessing, i.e., via modality translation, where the colored postfix means the target modality.
}
\vspace{-4mm}
\label{tab:statistics}%
\end{table*}%

\paragraph{Invocation-based Meta-response Tuning.}
To teach the LLM to produce the correct invocation meta responses, we need to create data specifically designed for instruction tuning. 
Technically, we make use of the existing annotated datasets for various vision tasks included in this work.
For each task under specific different user input scenarios, with the corresponding data, we design various template dialogue-format examples.
Based on these examples we then prompt the GPT-4 to generate more samples under various topics and enriched scenarios.

% \vspace{-2mm}
\section{A Benchmark for Grounded MUIE}

\vspace{-1mm}
To evaluate the performance of our grounded MUIE system, we develop a benchmark testing set.

\vspace{-1mm}
\subsection{Data Source}

We select 9 existing datasets from different modalities (or combinations thereof) for IE/MIE tasks. 
Table \ref{tab:statistics} summarizes these datasets of the raw sources.
We then process these datasets, such as Text$\leftrightarrow$Speech, to create 6 new datasets under new multimodal (combination) scenarios.
Before annotation, we carefully select 200 instances from their corresponding testing sets, ensuring each instance contained as much IE information as possible.

\begin{compactitem}
    \item \textbf{PASCAL-C}\footnote{\url{https://github.com/bethgelab/robust-detection-benchmark}}: is an object detection dataset for evaluating the robustness.  
    
    \item \textbf{VRD}\footnote{\url{https://cs.stanford.edu/people/ranjaykrishna/vrd/}}, Visual Relationship Dataset: is designed to assess the precision of detecting interactions among pairs of objects. Comprising 5,000 images, the dataset encompasses 100 object categories and 70 predicates.

    \item \textbf{imSitu}\footnote{\url{http://imsitu.org/}}: serves as a resource for facilitating situation recognition, a task concerned with generating a succinct depiction of the scenario portrayed in images.
    This includes (1) the main activity, (2) the participating actors, objects, substances, and locations and most importantly (3) the roles these participants play in the activity.

    \item \textbf{ACE2005}\footnote{\url{http://projects.ldc.upenn.edu/ace/}}: is extensively used in information extraction. It comprises annotated news articles in English, covering a diverse range of topics and events, with annotations including named entities, relations, and events.

    \item \textbf{ReTACRED}\footnote{\url{https://github.com/gstoica27/Re-TACRED}}: is a revised version of TACRED for relation detection, containing over 91k sentences spread across 40 relations.

    \item \textbf{VidSitu}\footnote{\url{https://vidsitu.org/}}: is a large-scale dataset containing diverse 10-second videos from movies depicting complex situations (a collection of related events). Events in the video are richly annotated at 2-second intervals with verbs, semantic-roles, entity co-references, and event relations.

    \item \textbf{Twt17}\footnote{\url{https://github.com/jefferyYu/UMT}}, Twitter-17: is a publicly available Twitter dataset for NER. It encompasses 723 test tweets, with annotations covering four entity types, namely, \textit{person}, \textit{location}, \textit{organization}, \textit{miscellaneous}.

    \item \textbf{MNRE}\footnote{\url{https://github.com/thecharm/MNREp}}, Multimodal Neural Relation Extraction: comprises 15,484 samples and 9,201 accompanying images across 23 distinct relation categories, partitioning into training, development, and testing subsets, consisting of 12,247, 1,624, and 1,614 samples respectively.

    \item \textbf{M$^2$E$^2$} \footnote{\url{https://github.com/limanling/m2e2}}: is comprised of 245 multimedia news articles meticulously annotated with events and their corresponding arguments.

\end{compactitem}

\begin{table*}[t]
  \centering
\fontsize{8}{10.5}\selectfont
\setlength{\tabcolsep}{1.9mm}
% \resizebox{0.99\textwidth}{!}{
\begin{tabular}{lcccccccccccccc}
\hline
		
\multicolumn{1}{c}{\multirow{3}{*}{\bf Method}} & \multicolumn{7}{c}{{\textbf{T+I Input}}} & \multicolumn{7}{c}{{\textbf{I Input}}} \\
\cmidrule(r){2-8}\cmidrule(r){9-15}
& \multicolumn{2}{c}{{\textbf{Twt17}}}
& \multicolumn{2}{c}{{\textbf{MNRE}}}
& \multicolumn{3}{c}{{\textbf{M$^2$E$^2$}}} 
& \multicolumn{2}{c}{{\textbf{PASCAL-C}}}
& \multicolumn{2}{c}{{\textbf{VRD}}}
& \multicolumn{3}{c}{{\textbf{imSitu}}} \\
% & \multicolumn{1}{c}{\multirow{2}{*}{\bf \emph{Avg}}}\\
\cmidrule(r){2-3}\cmidrule(r){4-5}\cmidrule(r){6-8}
\cmidrule(r){9-10}\cmidrule(r){11-12}\cmidrule(r){13-15}
& NER& I-Seg&  RE& I-Seg&  ET& EA& I-Seg&  NER& I-Seg& RE& I-Seg&  ET& EA& I-Seg \\

\hline
LLaVA\scriptsize{+SEEM}& 	23.0& 	45.8& 		15.4& 	51.8& 		22.8& 	13.5& 	48.3&  17.8&	26.1&		10.4&	36.9&	19.5&	8.2&	29.8\\
InstructBLIP\scriptsize{+SEEM}& 	26.9& 	52.0& 		17.0& 	54.8& 	 	23.0& 	20.1& 	52.5& 	   21.0& 	39.7& 		11.6& 	39.1& 		18.5& 	11.3& 	30.0 \\
MiniGPT-v2& 	45.4& 	48.7& 		22.4& 	56.2& 	27.3& 	16.3& 	54.8& 	 41.8&	62.0&		18.6&	38.3&		37.0&	13.3&	32.2\\
% \hline
\rowcolor{nmblue} \textsc{Reamo} & \bf 47.4& 	\bf 53.5& 	 	\bf 24.6& 	\bf 56.9& 	 	\bf 30.2& 	\bf 25.6& 	\bf 60.1& 	   \bf 43.0& 	\bf 64.6& 	 	\bf 26.0& 	\bf 43.9& 	\bf 41.5& 	\bf 16.3& 	\bf 39.6 \\

\hline

\end{tabular}
% \vspace{-2mm}
\caption{
Zero-shot performance in the UIE scenario of text+image or standalone image input.
I-Seg: grounding by image segmentation. 
Performance of \textsc{Reamo} is with blue background.
}
% \vspace{-4mm}
  \label{tab:main-T-I}%
\end{table*}%

\paragraph{Pre-processing And Modality Translation}

Before we begin the annotation work for grounding, we enrich the types of modality combinations. 
This is essential because most MIE datasets currently focus on the combination of images and text. 
However, we aim to simulate a variety of common modality combinations that could occur in real-world scenarios. 
To achieve this, we transform and preprocess existing datasets. 
Specifically, our approach involves cross-modal parallel translation to generate data in another modality. 
We preprocess the following datasets:

\begin{compactitem}

    \item \textbf{VidSitu\textcolor{deepgreen}{-Aud}}: we start by captioning videos from the VidSitu dataset, then use the given video event annotations combined with the captions to have ChatGPT generate a coherent sentence, serving as the paired text for each piece of data.

    \item \textbf{VidSitu\textcolor{deepred}{-Aud}}: Based on VidSitu-Txt, we convert each sentence into speech using Text-To-Speech (TTS) tools. We employ state-of-the-art open-source TTS models: Bark\footnote{\url{https://github.com/suno-ai/bark}} and Edge-TTS\footnote{\url{https://github.com/rany2/edge-tts}}.

    \item \textbf{ACE\textcolor{deepred}{-Aud}}: we take original textual sentences from the ACE dataset and record speech using TTS technology.

    \item \textbf{MNRE\textcolor{deepred}{-Aud}}: we record speech for sentences from the MNRE using TTS.

    \item \textbf{Twt17\textcolor{deepred}{-Aud}}: Similarly, we record speech for sentences from the Twt17 dataset using TTS.
    
\end{compactitem}

However, we emphasize that such parallel data generation can only produce modality-aligned content. 
To create diverse content, we plan to introduce randomness by adding noise to some instances. For example, we might alter parts of the original text before synthesizing the speech with TTS.
Considering the aim to only annotate a test set for the sake of conserving labor and reducing costs, we plan to select subsets from the test sets of various original datasets obtained. We aim to select 200 from each. 
Our selection criterion focuses on ensuring a high quantity of entities and objects in the content, and the final labels should cover a rich vocabulary. 
Given there are 15 combinations of modalities and tasks (including those augmented through our post-processing), we will have a total of 3,000 data entries of grounded MUIE test instances.
Next, we revisit the annotation information, specifically re-annotating instances from the combined modality datasets where cross-modal content is not fully aligned. 
This ensures that the dataset covers both modality-shared and modality-specific instances.

\vspace{-1mm}
\section{Experiments}

\vspace{-2mm}
\subsection{Settings}
\vspace{-1mm}

We measure the system performance by following most practices of end-to-end UIE: F1 of entity span with type for NER,
F1 of all subject\&object entities and their relation label for RE.
For EE, we consider event trigger (ET) F1 including both trigger and event type, and event argument F1 including both arguments with role types.
W.r.t. the multimodal grounding, for both image and audio segmentation, we consider the mean Intersection over Union (mIoU); for video segmentation, we use the average Jaccard (J).

\begin{table*}[t]
  \centering
\fontsize{8}{10}\selectfont
\setlength{\tabcolsep}{4mm}
\begin{tabular}{lcccccccc}
\hline

\multicolumn{1}{c}{\multirow{2}{*}{\bf Method}} & \multicolumn{4}{c}{{\textbf{T+A Input}}} & \multicolumn{4}{c}{{\textbf{A Input}}} \\
\cmidrule(r){2-5}\cmidrule(r){6-9}
& \multicolumn{2}{c}{{\textbf{ACE05-Aud}}}
& \multicolumn{2}{c}{{\textbf{ReTACRED}}}
& \multicolumn{2}{c}{{\textbf{ACE05-Aud}}}
& \multicolumn{2}{c}{{\textbf{ReTACRED}}}  \\
\cmidrule(r){2-3}\cmidrule(r){4-5}\cmidrule(r){6-7}\cmidrule(r){8-9}
& NER& A-Seg& RE& A-Seg& NER& A-Seg& RE& A-Seg \\

\hline

% \hline
SpeechGPT& 	26.7& 	21.4& 	45.4& 	27.5& 	14.0& 	13.3& 	30.4& 	21.0 \\
NExT-GPT\scriptsize{+SHAS}& 19.6& 	15.6& 		37.5& 	20.4&    8.3& 	10.2& 	25.1& 	12.4 \\

\rowcolor{nmblue} \textsc{Reamo} & 	\bf 28.5& 	\bf 24.3& 	\bf 46.8& \bf 29.1&  \bf  17.4	& \bf 16.7&  	\bf 33.4& 	\bf 25.1 \\

\hline
\end{tabular}
% \vspace{-2mm}
\caption{
Zero-shot performance in the text+audio or standalone audio input scenarios.
}
% \vspace{-3mm}
  \label{tab:main-T-A}%
\end{table*}%

\begin{table}[t]
  \centering
\fontsize{8}{10}\selectfont
\setlength{\tabcolsep}{1.2mm}
\begin{tabular}{lcccccc}
\hline
\multicolumn{1}{c}{\multirow{2}{*}{\bf Method}}
& \multicolumn{3}{c}{{\textbf{T+V (VidSitu-Txt)}} }
& \multicolumn{3}{c}{{\textbf{ V (VidSitu)}}} \\
\cmidrule(r){2-4}\cmidrule(r){5-7}
& ET& ER& V-Trck& ET& ER& V-Trck \\
\hline
VideoChat\scriptsize{+SEEM}& 	28.8& 	18.5& 	28.1& 		14.3& 	9.2& 	20.9 \\
Video-LLaVA\scriptsize{+SEEM}& 	31.0& 	22.4& 	31.4& 		18.6& 	8.8& 	20.6 \\

\rowcolor{nmblue}  \textsc{Reamo} & 	\bf 32.8& 	\bf 23.1& 	\bf 34.4& \bf 22.3& 	\bf 14.5& 	\bf 23.2 \\

\hline
\end{tabular}

\caption{
Zero-shot results in the text+video or standalone video input scenarios.
}
\vspace{-3mm}
  \label{tab:main-T-V}%
\end{table}%

The encoder projection is a linear layer with a hidden size of 4,096. 
As no prior method is designed for grounded MUIE, we implement pipeline systems.
Specifically, we first employ an MLLM to do UIE on multimodal input.
And then we pass the raw multimodal source and the necessary UIE label to the SEEM or SHAS model for fine-grained grounding of image, video and audio.
We consider the following existing well-exposed MLLMs.
For image-related UIE:
InstructBLIP \cite{abs-2305-06500},
LLaVA \cite{abs-2304-08485}, and also MiniGPT-v2 \cite{chen2023minigpt} that can output image segmentation end-to-end.
For or video-related UIE:
VideoChat \cite{abs-2305-06355},
Video-LLaVA \cite{abs-2311-10122}.
SpeechGPT \cite{abs-2305-11000} for audio-related UIE.
Video-LLaMA \cite{abs-2306-02858} supporting video+audio; NExT-GPT \cite{wu2023next} supporting all four modalities.
All these systems take the 7B LLM, unless otherwise specified.
For fairness, all baselines are further tuned using the same UIE instruction-tuning data as ours.
All system takes zero-shot inference, without tuning on in-house datasets.

\begin{table*}[!t]
  \centering
\fontsize{8}{10}\selectfont
\setlength{\tabcolsep}{2.5mm}
\begin{tabular}{lcccccccccc}
\hline

\multicolumn{1}{c}{\multirow{2}{*}{\bf Method}} & \multicolumn{3}{c}{{\textbf{T+I+A (Twt17-Aud)}}} & \multicolumn{3}{c}{{\textbf{I+A (MNRE-Aud)}}}  & \multicolumn{4}{c}{{\textbf{V+A (VidSitu-Aud)}}} \\
\cmidrule(r){2-4}\cmidrule(r){5-7}\cmidrule(r){8-11}
& NER& I-Seg& A-Seg& RE&I-Seg& A-Seg&  ET & ER& V-Trck& A-Seg \\

\hline
Video-LLaMA\scriptsize{+SEEM/+SHAS} & -& 	-& 	-& 	-& -& - &  	12.0& 	4.8	& 12.7	& 8.4\\

NExT-GPT\scriptsize{+SEEM/+SHAS} &	30.7 &	32.4 &	13.9 &	15.4 &	46.5 &	18.8  &	19.3 &	13.7 &	19.9 &	15.0 \\

\rowcolor{nmblue} \textsc{Reamo} & \bf 37.4 &	\bf 33.3 &	\bf 15.1  &	\bf 21.8 &	\bf 53.4 &	\bf 21.8 &	\bf 24.2 &	\bf 18.5 &	\bf 22.0 &	\bf 20.9 \\

\hline

\end{tabular}
% \vspace{-2mm}
\caption{
Zero-shot performance in more complex modality-hybrid scenarios of MUIE.
}
\vspace{-4mm}
  \label{tab:main-Complex}%
\end{table*}%

\vspace{-2mm}
\subsection{Zero-shot Results on Image-related MUIE}

\vspace{-2mm}
Table \ref{tab:main-T-I} presents the results of different models on our MUIE dataset under both text+image and pure image conditions. 
From the data, we observe that overall, the end-to-end approach (MiniGPT-v2) outperforms pipeline methods.
Besides, our system demonstrates a clear and consistent advantage consistently across all subtasks.

\begin{figure}[!t]
\centering
\includegraphics[width=1\columnwidth]{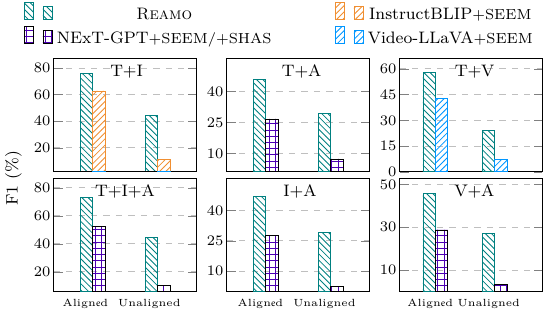}
% \vspace{-2mm}
\caption{
Performance gap between modality-shared (aligned) and modality-specific (unaligned) MUIE.
}
\label{fig:modal-align}
\vspace{-3mm}
\end{figure}

% \vspace{-2mm}
\subsection{Zero-shot Results on Audio-related MUIE}

% \vspace{-1mm}
In Table \ref{tab:main-T-A}, we present the performance of various models under text+audio and pure audio input, respectively. 
The trends observed are similar to those seen in the previous table:
1) End-to-end approaches (SpeechGPT) demonstrate stronger performance compared to the pipeline method (NExT-GPT+SHAS), effectively mitigating the issue of error propagation.
2) Our \textsc{Reamo} consistently outperforms others across all subtasks and scenarios.

% \vspace{-2mm}
\subsection{Zero-shot Results on Video-related MUIE}

% \vspace{-1mm}
In Table \ref{tab:main-T-V}, we present the final set of results for EE task based on text+video and pure video input. 
The overall trend observed here again aligns with that of the previous tables, with our model achieving the strongest performance. 
Following, we can confirm that for single modalities (such as image, audio, video), without the assistance of textual modality, the effectiveness of MUIE significantly diminishes.

\begin{figure}[!t]
    \centering
    \includegraphics[width=1\columnwidth]{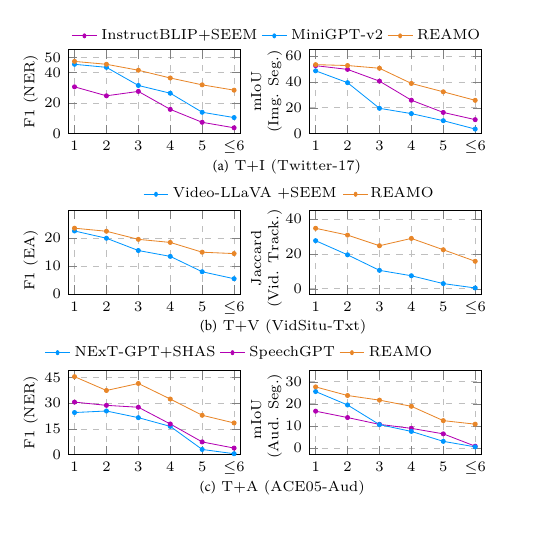}
    % \vspace{-2mm}
    \caption{
    Impact of different object/entity numbers. 
    }
    \label{fig:entity-num}
    \vspace{-1mm}
\end{figure}

\vspace{-1mm}
\subsection{Results on Modality-Compound MUIE}

\vspace{-1mm}
Finally, we present the performance of our model across a broader range of complex mixed-modality scenarios for MUIE. 
As shown in Table \ref{tab:main-Complex}, our \textsc{Reamo} consistently delivers the best results than the baseline solutions. 
This becomes even more pronounced in complex mixed-modality contexts.

\vspace{-1mm}
\subsection{Gap between Modality-Shared vs. Modality-Specific MUIE}

\vspace{-1mm}
Previously, we mentioned that in the MUIE context, different models may encounter both aligned shared information and misalignments between modalities. 
Now, we have categorized cases in the data as either Modality-Shared or Modality-Specific and compared the performance of various models under these two scenarios. 
As illustrated in Fig. \ref{fig:modal-align}, \textsc{Reamo} outperforms all baseline models in addressing challenges arising from misalignments between modalities.

\begin{figure}[!t]
\centering
\includegraphics[width=0.98\columnwidth]{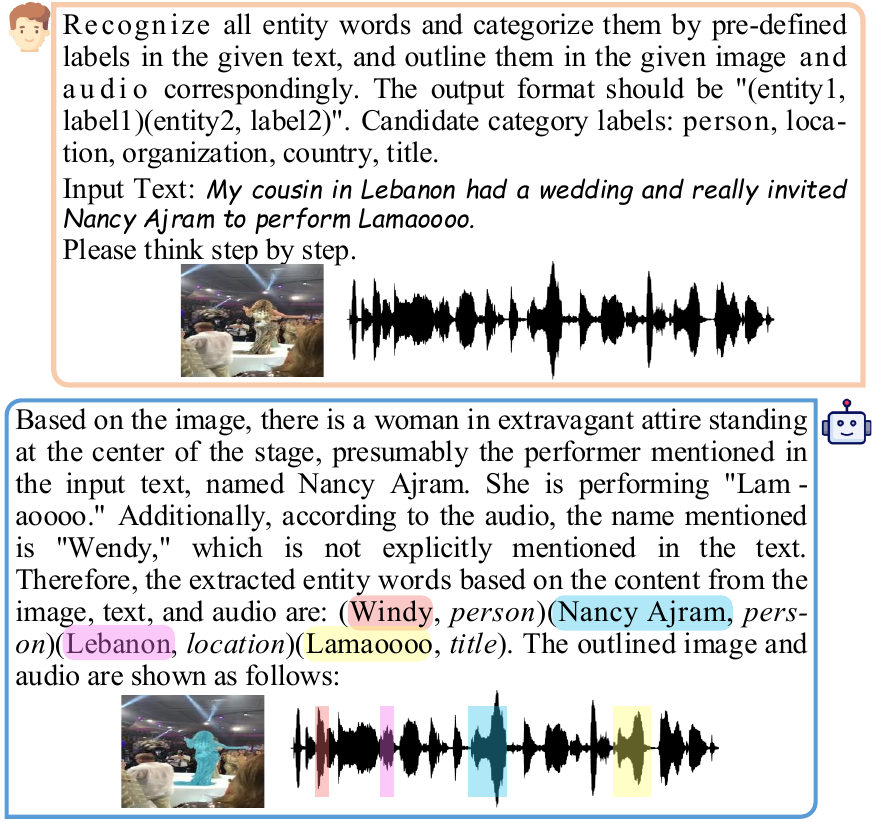}
\vspace{-3mm}
\caption{
Qualitative result A on MUIE (NER) with modality-specific case via reasoning.
}
\label{fig:case1}
% \vspace{-3mm}
\end{figure}

\begin{figure}[!t]
\centering
\includegraphics[width=0.98\columnwidth]{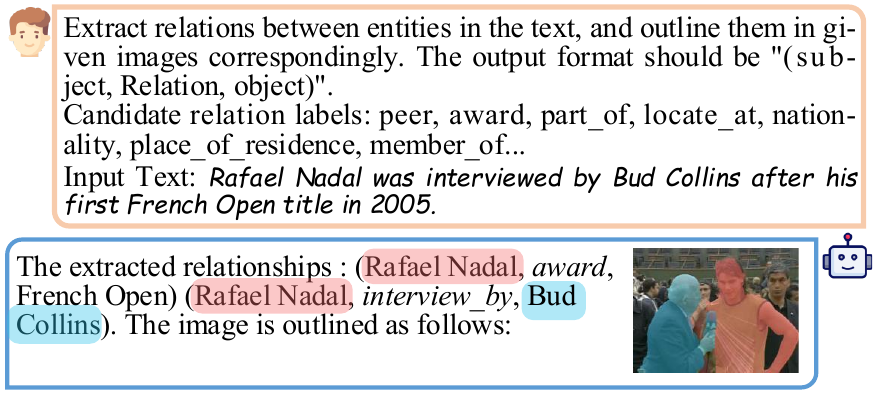}
\vspace{-2mm}
\caption{
Qualitative result B on MUIE (RE) with grounding rationale via reasoning.
}
\label{fig:case2}
\vspace{-3mm}
\end{figure}

\vspace{-2mm}
\subsection{Influence of Entity/Object Numbers}

\vspace{-1mm}
Fig. \ref{fig:entity-num} further illustrates the impact of the number of entities (or objects) in an instance on the performance of the MUIE system. 
It is evident that our \textsc{Reamo} system maintains commendable performance in extracting an increasing number of objects across T+I/T+A/T+V scenarios, both in terms of MUIE results and fine-grained grounding, being clearly superior to baseline pipeline systems.

% \vspace{-2mm}
\subsection{Case Study}

\begin{figure}[!t]
\centering
\includegraphics[width=0.98\columnwidth]{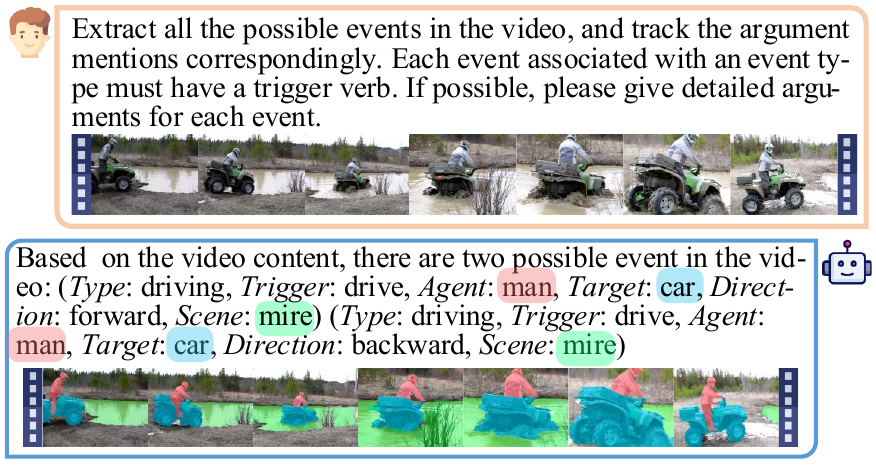}
\vspace{-2mm}
\caption{
Qualitative result C on MUIE (EE) with commonsense-aware cognitive reasoning.
}
\label{fig:case3}
\vspace{-3mm}
\end{figure}

% \vspace{-1mm}
Then, we provide visualizations of case studies, through which we aim to offer a more intuitive and comprehensive demonstration of our MUIE system. 
Fig. \ref{fig:case1}, \ref{fig:case2}, and \ref{fig:case3} each displays an example for NER, RE, and EE tasks, respectively, where \textsc{Reamo} yields correct answers.
In case A, our system demonstrates how it can engage in thought processes and flexibly determine MUIE labels from different modalities.
In case B, the model can provide accurate relation extractions and precise segmentation results that correspond to the entities in the input text.
In case C, most impressively, it manages to link semantic information to factual commonsense understanding based on accurate fine-grained cross-modal grounding, and correctly answer the question.

\subsection{Error Analysis}
Finally, we delve into our model itself, and analyze the possible shortcomings for shedding light on future explorations.
Via our experiments, we here summarize several error types:
\begin{compactitem}
    
    \item \textbf{Repetition of Extracted Content}: When text and information from other modalities are not strictly consistent, our method may output different entity names, arguments, or relationships. However, upon integrating information from different modalities, they should correspond to the same entity names, arguments, or relationships.
    \item \textbf{Incomplete Information Extraction}: The outcomes of information extraction are incomplete, such as incomplete named entity recognition, failure to identify relations involving in-depth reasoning, or incomplete identification of event arguments.
    \item \textbf{Incorrect Grounding Match}: The entity or arguments do not always match with the grounding results. 
    For instance, when the text mentions `Obama' and `Trump' and the image depicts both individuals, the image object segmenter fails to ascertain who is `Obama' and `Trump', resulting in an erroneous grounding match.
    \item  \textbf{Miss-grounding}: Our model may output entities or arguments without successfully grounding the corresponding regions in the respective image, video, or audio.
    \item \textbf{Over-grounding}: The model may generate multiple instructions and perform grounding in the image, video, or audio, yet no corresponding regions actually exist in the visual or auditory content.
    \item \textbf{Error Propagation}: Since our system operates as a pipeline process, where a meta-response is first produced and then used to invoke functional modules, this sequence introduces error propagation. If there is an issue with the content of the meta-response, the outcomes from the subsequent modules will be incorrect. To address this, we need to develop more advanced end-to-end MLLMs.
    
\end{compactitem}

\vspace{-1mm}
\section{Conclusion}

\vspace{-2mm}

This paper introduces a novel task of multimodal information extraction setting: grounded Multimodal Universal Information Extraction (MUIE).
First, MUIE definition unifies all IE tasks across various modalities, including text, audio, images, and video, with fine-grained multimodally grounded targets. 
To solve MUIE, we devise \textsc{Reamo}, a novel MLLM that can extract and ground information from all modalities. 
\textsc{Reamo} is tuned through various strategies to achieve proficiency in recognizing and grounding multimodal information. 
Further we introduce a high-quality, diverse, and challenging benchmark dataset for evaluating MUIE systems.
Experimental results demonstrate \textsc{Reamo}'s stronger performance in extracting and grounding, setting a strong benchmark for the following grounded MUIE research.

\vspace{-1mm}
\section*{Acknowledgments}

\vspace{-2mm}
This work is supported by the CCF-Baidu Open Fund, 
and the Research on automatic construction of cross-domain knowledge graph with multi-modal information (No. U23B2055)

\vspace{-1mm}
\section*{Limitations}

\vspace{-2mm}
The limitations of this work primarily arise from the following two aspects.
From model perspective, our MLLM, designed for Grounded MUIE, handles four common modalities (text, images, video, audio) and may struggle to integrate new ones without significant retraining. Challenges persist in extracting complex implicit information and in cognitive reasoning, with current capabilities mainly supporting video tracking for event extraction. 
Further research is needed to enhance and expand tracking for other tasks like NER and RE in videos.

From data perspective, we introduced a dataset solely comprising a test set, limiting in-house training opportunities. Future work will focus on enlarging this dataset to include a training set and expanding annotations to cover more modality combinations for the three IE tasks: NER, RE, and EE.

\vspace{-1mm}
\section*{Ethics Statement}

\vspace{-2mm}

The development and application of MUIE systems potentially raise several potential ethical considerations or risks that should be properly treated to ensure responsible research and deployment.

\vspace{-2mm}
\paragraph{Privacy and Consent.}
Multimodal data in MUIE systems can include sensitive personal information. Ensuring data is collected and used with explicit consent and in compliance with data protection laws is critical.

\vspace{-2mm}
\paragraph{Bias and Fairness.}
MLLMs in MUIE systems may inherit biases from their training data, potentially causing discriminatory effects. Active efforts are required to mitigate such biases to guarantee fairness across diverse groups.

\vspace{-2mm}
\paragraph{Transparency and Accountability.}
MUIE systems often lack clear decision-making transparency due to their complexity. Promoting explainable AI with detailed documentation can enhance user and stakeholder trust.

\vspace{-2mm}
\paragraph{Misuse Potential.}
The ability of MUIE systems to process diverse data can be exploited for harmful purposes like misinformation or unauthorized surveillance. Establishing robust legal and ethical safeguards is essential to prevent misuse.

\bibliography{ref}

\clearpage

\appendix

\section{Model Prompt Specification}
\label{Model Specification}

Here, we show the prompts for each subtask (i.e., NER, RE, EE) in MUIE:

\begin{tcolorbox}[fontupper=\linespread{0.9}\selectfont,]
{
\small
$\blacktriangleright$ \textbf{Prompt for Named Entity Recognition:}\\
Please recognize all entity words and categorize them by pre-defined labels in the given text, and outline them in the given image or video or audio correspondingly. The output format should be ``(entity1, label1)(entity2, label2)''.
If an entity possibly has a counterpart in the given image or video or audio, please generate a token ``<concept>'' after the entity word, for subsequent cross-modal grounding.\\
Candidate category labels: \textit{person, location, organization, country...}\\
\textbf{Input Text/Image/Video/Audio:} \textlogo \imagelogo \videologo \audiologo.
}
\end{tcolorbox}

\begin{tcolorbox}[fontupper=\linespread{0.9}\selectfont,]
{
\small
$\blacktriangleright$ \textbf{Prompt for Relation Extraction:}\\
Please extract all relations between named entities, and outline them in the given image or video or audio correspondingly. The output format should be ``(subject entity, relation, object entity)''.
If an entity possibly has a counterpart in the given image or video or audio, please generate a token ``<concept>'' after the entity word, for subsequent cross-modal grounding.\\
Candidate relation labels: \textit{peer, award, part\_of, locate\_at, nationality, place\_of\_residence, member\_of...}\\
\textbf{Input Text/Image/Video/Audio:} \textlogo \imagelogo \videologo \audiologo.
}
\end{tcolorbox}

\begin{tcolorbox}[fontupper=\linespread{0.9}\selectfont,]
{
\small
$\blacktriangleright$ \textbf{Prompt for Event Extraction:}\\
Extract all the possible events in the video, and track the argument mentions correspondingly. Each event associated with an event type must have a trigger verb. If possible, please give detailed arguments for each event.\\
Candidate event types: \textit{Marry, Attack, Injure, Be-born, Meet, Transport, Start-position...}, \\
Candidate event argument types: \textit{Agent, Target, Direction, Time, Place, Instrument, Organization, Duration... }\\
\textbf{Input Text/Image/Video/Audio:} \textlogo \imagelogo \videologo \audiologo.
}
\end{tcolorbox}

\section{Specification of MUIE Fine-tuning}
\label{Specification of MUIE Fine-tuning}

\subsection{UIE Instruction Tuning}
\label{UIE Instruction Tuning}

In this step, we train only the core LLM. To avoid the significant cost associated with fully updating the LLM, we employ the LoRA technique, which allows us to achieve our objectives by tuning only a small subset of parameters, thus leaving the overall architecture of the LLM unchanged. The datasets we utilize are sourced from existing IE datasets. We list these datasets in Table \ref{tab:UIE-tuning-data}. These data are converted into an instruction format, following the practices outlined by \citet{wang2023instructuie}.

\begin{table}[h]
 \centering
\fontsize{9}{12}\selectfont
\setlength{\tabcolsep}{2mm}
% \resizebox{0.98\columnwidth}{!}{
\begin{tabular}{llc}
\hline
\multicolumn{1}{c}{\multirow{1}{*}{\textbf{Task}}}
 & \multicolumn{1}{c}{ \textbf{Data Source}} 
 & \multicolumn{1}{c}{ \textbf{Amount}} \\
\hline

\multirow{1}{*}{\textbf{NER}} &  OntoNotes 5.0\footnote{\url{https://catalog.ldc.upenn.edu/LDC2013T19}}& 76,714\\
 & CoNLL2003\footnote{\url{https://huggingface.co/datasets/conll2003}} &  20,744\\
 
 \hline
\multirow{1}{*}{\textbf{RE}} &  NYT \cite{RiedelYM10} & 56,196\\
& NYT11 HRL \cite{TakanobuZLH19} &  62,648\\
 \hline
\multirow{1}{*}{\textbf{EE}} & MAVEN \cite{WangWHJHLLLLZ20}	& 49,873\\

\hline
\end{tabular}
% \vspace{-2mm}
\caption{
Datasets used for UIE tuning.
}
% \vspace{-4mm}
\label{tab:UIE-tuning-data}%
\end{table}%

\subsection{Multimodal Alignment Learning}

To accomplish the alignment, we adopt an `X-to-text' generation task trained on the `X-caption' pair (`X' stands for image, audio, or video) data from existing corpus and benchmarks, i.e., given the representation of an `X', to prompt the frozen LLM to generate the corresponding text description.
Specifically, we utilize three types of `X-caption' pair data, including: 
1) `Video-caption' pair dataset: Webvid-2M \citep{BainNVZ21}, a large-scale dataset of short videos with textual description sourced from stock footage sites, 
2) `Image-caption' pair dataset: CC3M \citep{SoricutDSG18}, contains over 3 million images accompanied by diverse styles of natural-language descriptions, 
and 3) `Audio-caption' pair dataset: AudioCaps \citep{KimKLK19}, an extensive dataset of approximately 46k audio clips paired with human-written textual descriptions collected via crowdsourcing.

\subsection{Fine-grained Cross-modal Grounding-aware Tuning}

For this step, our focus is on achieving fine-grained, concept-level cross-modal alignment, primarily aligning the other three modalities to the textual modality. 
Our primary method involves utilizing existing `X-to-text' phrase grounding datasets. 
\textsc{Reamo}'s input comprises textual phrases and grounded regional modality features, prompting the LLM to determine their match. 
For fine-grained image-to-text alignment, we consider the MS-COCO dataset \cite{lin2014microsoft}. 
For video-to-text alignment, we turn to the TAO dataset \cite{DaveKTSR20}. 
For entity-level audio-to-text alignment, we primarily utilize the dataset for the Speech NER task \cite{ChenXWXZH22}.
Since there is a lack of dataset in English speech, we thus use TTS tools to generate the CoNLL 2003 textual NER data into speech NER data.

\paragraph{Invocation-based Meta-response Tuning.}

For this step, we aim to teach the LLM to produce the correct format of invocation meta-responses.
Thus, we need to create data specifically designed for instruction tuning. 
Technically, we make use of the existing annotated datasets for various vision tasks included in this work.
For each task under specific different user input scenarios, with the corresponding data, we design various template dialogue-format examples.
Based on these examples we then prompt the GPT-4 to generate more samples under various topics and enriched scenarios.
To ensure data diversification, the resulting instruction data includes a rich number of instances (10k), under different simulations of modalities and tasks.

\section{Grounded MUIE Evaluation}
\label{sec:appendixC}

Our grounded MUIE evaluation dataset involves predictions for three tasks, including UIE label prediction, multimodal grounding prediction, and cognitive QA task prediction. 
Here, we provide detailed evaluation metrics for these three subtasks.

\subsection{UIE Evaluation Metrics}
To evaluate textual UIE results of the model, we use \texttt{span-based offset Micro-F1} as the primary
metric.
\begin{compactitem}
    \item For NER task, we follow a span-level evaluation setting, where the entity boundary and entity type must be correctly predicted.
    \item For RE task, a relation triple is correct if the model correctly predicts the boundaries of the subject entity, the object entity, and the entity relation.
    \item For EE task, we report two evaluation
metrics: 
    \begin{compactitem}
        \item  Event Trigger (ET): an event trigger is
    correct if the event type and the trigger word are
    correctly predicted.
        \item Event Argument (EA): an event argument is correct if its role type and event type match a reference argument mention.
    \end{compactitem}
\end{compactitem}

\subsection{Modality Grounding Evaluation Metrics}
For the evaluation of the fine-grained modality grounding accuracy, the key idea is to measure the \texttt{mean Intersection over Union (mIoU)}.

\paragraph{Image Segmentation.}
Let us denote by $\hat{M}_{img} = \{M_g\}^G_{g=1}$ the ground truth set of $G$ regions, and ${M}_{img} = \{M_k\}^K_{k=1}$ the set of $K$ predictions. 
Inspired by prior work, if $K \neq G$, we employ padding with $\emptyset$ to equalize the sizes of both sets, resulting in a final size of $P =\operatorname{max}(G, K)$.
Then, we find a bipartite matching between these two sets by
searching for a permutation of $P$ elements, $\sigma \in \mathcal{S}_P$, with the lowest cost:
\begin{equation}
    \hat{\sigma} = \operatorname{arg min}_{\sigma \in \mathcal{S}_P} \sum_{i}^P \mathcal{L}_{match}(\hat{M}_i, M_{\sigma(i)}),
\end{equation}
where $\mathcal{L}_{match}(\hat{M}_i, M_{\sigma(i)})$ is a pairwise matching cost between ground truth Mi and a prediction with index $\sigma(i)$.
We compute this optimal assignment efficiently with the Hungarian algorithm.
 We define $\mathcal{L}_{match}(\hat{M}_i, M_{\sigma(i)})$ as
$\mathcal{L}_{bce}(\hat{M}_i
, M_{\sigma(i)}) + \mathcal{L}_{dice}(\hat{M}_i
, M_{\sigma(i)})$. 
The final IoU of each prediction is:
\begin{equation}
    \text{IoU} = \frac{\text{Area of Overlap}}{\text{Area of Union}}
\end{equation}
Based on the IoU scores, we can calculate mIoU metric by referring image segmentation dataset.

\paragraph{Video Tracking.}

For videos, we compute the Jaccard Index (a.k.a, mIoU score) for each frame via the above calculations, and then average them.

\paragraph{Audio Segmentation.}

Similarly, the mIoU score for each audio segment is computed to evaluate the quality of speech segmentation results.
We measure the 1D span of the extracted segments and the 1D span of gold segments.

\end{document}